%% file: main.tex
\documentclass[lettersize,journal]{IEEEtran}
\usepackage{amsmath,amsfonts}
\usepackage{algorithmic}
\usepackage{array}
\usepackage[caption=false,font=normalsize,labelfont=sf,textfont=sf]{subfig}
\usepackage{textcomp}
\usepackage{stfloats}
\usepackage{url}
\usepackage{verbatim}
\usepackage{graphicx}
\usepackage{enumitem}
\def\BibTeX{{\rm B\kern-.05em{\sc i\kern-.025em b}\kern-.08em
    T\kern-.1667em\lower.7ex\hbox{E}\kern-.125emX}}
\usepackage{cite}
\usepackage{balance}
\usepackage{booktabs}
\usepackage{tabularx}
\usepackage{multirow}
\usepackage{xcolor}
\usepackage{amsmath}

\title{MSCENet: A Multi-Scale Correlation Enhanced Network for Anomaly Detection}

\author{Long Zhao†,
	    Shixun Ji†,
        Zhipeng Wang, \IEEEmembership{Member, IEEE,}
       Bin Cheng, \IEEEmembership{Member, IEEE,}
        and Bin He, \IEEEmembership{Senior Member, IEEE}
    \thanks{†These authors contributed equally to this work.}
    \thanks{This work was supported in part by the National Natural Science Foundation of China under grants 62573322 and 62088101, by Shanghai Rising-Star Program under grant 24QA2709400, by the Shanghai Chenguang Program under grant 22CGA19, and by the Shanghai Municipal Science and Technology Major Project under grant 2021SHZDZX0100. (Corresponding author: Bin Cheng.)}
    \thanks{The authors are with the Department of Control Science \& Engineering, Tongji University, Shanghai 201804, China, and also with the National Key Laboratory of Autonomous Intelligent Unmanned Systems, Shanghai 201203, China (e-mail: superzhaolong@tongji.edu.cn; jsx@tongji.edu.cn; wangzhipeng@tongji.edu.cn; bincheng@tongji.edu.cn; hebin@tongji.edu.cn).}
}

\begin{document}

\maketitle

\begin{abstract}
In the field of multivariate time series anomaly detection, against the backdrop of increasing data complexity and complex dependencies across multiple temporal scales, traditional methods often struggle to simultaneously capture temporal dynamic features and intricate inter-series correlations. To address this, we propose an innovative framework, MSCENet, which leverages advanced spatio-temporal learning and multi-scale learning techniques to enhance detection accuracy. MSCENet includes a fine-grained temporal convolution module that captures complex temporal dependencies through dilated convolutions, enabling the detection of both short- and long-term patterns. Additionally, the framework models inter-series relationships as a graph structure, using Mixhop graph convolutions to adaptively capture spatial dependencies across varying time scales. To support robust anomaly detection, the multi-scale gated convolution module in MSCENet integrates spatial and temporal attributes through gated mechanisms, facilitating the detection of subtle variations across multiple scales. Experimental evaluations on real-world datasets—SMD, PSM, and SWaT. It provides an adaptable and high-performance solution for anomaly detection in complex time series data environments.
\end{abstract}

\begin{IEEEkeywords}
Time series analysis, Multi-scale correlation, Anomaly detection, Machine learning, Deep learning.
\end{IEEEkeywords}

\section{Introduction}
\label{sec:intruduction}
\IEEEPARstart{T}{ime} series data consists of observations collected at regular intervals, often exhibiting strong temporal dependencies and recurring patterns. It is widely used across sectors such as manufacturing, healthcare, transportation, and network environments, where it includes sensor measurements, network traffic, vehicle motion, and medical information \cite{liu2024multivariate, Vu2023}. Time series analysis typically employs regression techniques for forecasting and classification methods for anomaly detection \cite{schmidl2022anomaly}, each serving distinct functions. Point anomalies, contextual anomalies, and collective anomalies represent the three primary categories of anomalies in time series data. Point anomalies refer to individual data points that show a marked deviation from the majority of the data. Observations that differ from expected patterns in a given temporal context, often identified through sliding windows, are known as contextual anomalies. On the other hand, collective anomalies occur when clusters of related data points show abnormal behavior in relation to the overall dataset. The variety of anomaly patterns creates significant obstacles for the reliable detection of irregularities within time series data.

As time series data becomes more intricate, particularly with the rise of industrial IoT, dependencies within multivariate time series are magnified \cite{Zeng2023}, \cite{Li2024}, complicating the task of anomaly detection. The added reliance on labeled data makes unsupervised anomaly detection even more difficult. Early approaches, such as clustering \cite{Scholkopf2001}, density-based \cite{Breunig2000a}, and shape-based methods \cite{Feng2021}, are limited in scalability and accuracy for high-dimensional, complex time series. Modern approaches are typically divided into predictive and reconstructive methods, both of which enhance detection accuracy by modeling normal data dependencies and calculating anomaly scores through comparisons between observed and expected behaviors, either predicted or reconstructed from time series dependencies \cite{Deng2021, Zhang2024}. High anomaly scores typically highlight events that require further scrutiny. While prediction-based methods excel at capturing temporal dependencies and adapting to shifts in time series dynamics, their reliance on historical data makes them vulnerable to unseen patterns, reducing their robustness in certain contexts. In contrast, methods based on reconstruction present a fuller representation of the data and take advantage of advanced non-linear modeling capabilities \cite{Jin2024}. Multivariate anomaly detection, which focuses on capturing interdependencies between multiple sequences, presents an additional challenge. Anomalies in one sequence can be influenced by patterns in others, complicating the detection process.

However, while these modern approaches represent a significant improvement, they often overlook the diverse relationships that span multiple time scales. This limitation is significant for accurate anomaly detection, as it prevents models from capturing critical variations in inter-series dependencies. For instance, physiological signals (e.g., heart rate, blood pressure) exhibit dynamic correlations during medical emergencies, and industrial sensor readings show fluctuating inter-dependencies during anomalies. An effective technique for modeling inter-series correlations is to treat multivariate time series as graphs. Sensors are modeled as nodes in a graph structure, where the features of each node adjust over time to reflect the system's state changes. Relying on only one kind of inter-series correlation, for instance, employing graph neural networks (GNNs) where the graph structure remains unchanged \cite{HAN2024103630}, can lead to reduced anomaly detection accuracy and suboptimal effectiveness in situations defined by complicated and fluctuating inter-series correlations. Though certain approaches take into account evolving graph structures \cite{LI2023119374}, they frequently overlook that these correlations are often closely linked to consistent time scales, such as economic and environmental cycles. This oversight can result in overlooked anomalies and increased false alarms in detection.

To overcome the shortcomings of earlier methods, we propose an innovative multi-scale correlation enhanced network (MSCENet) algorithm designed to detect anomalies within multivariate time series. Our method represents multivariate time series data as a graph, integrating a GNN to capture inter-series relationships and a convolutional neural network to extract temporal features. Anomaly scores are calculated by comparing observed values with forecasted values from the model, enabling the detection of anomalies \cite{Li_Li_Zhang_Wu_2019}. In more detail, the fine-grained temporal convolutional neural networks module enhances the model's capacity to capture complex temporal dependencies within time series data by utilizing dilated convolutions, which expand the receptive field without increasing the parameter count. The spatial correlation learning through graph structures module focuses on capturing inter-series relationships by modeling time series data as a graph structure. Leveraging a Mixhop graph convolution approach, this module dynamically learns spatial dependencies between variables, enabling the model to adapt to fluctuating correlations across multiple time scales. Finally, the multi-scale gated convolution for spatial-temporal feature extraction module further refines feature extraction by applying multi-scale convolutions with gated mechanisms, effectively integrating spatial and temporal features for enhanced anomaly detection. We outline our contributions in three key points:
\begin{itemize}
  \item We introduce MSCENet, an advanced framework for anomaly detection in multivariate time series. It separately captures temporal and feature dependencies at varying scales, utilizing correlation learning within a spatial-temporal network to ensure efficient and accurate detection of anomalies by modeling both temporal dynamics and scale interactions across time.
  \item We design and integrate specialized modules within MSCENet, including a fine-grained temporal convolutional neural networks module, a spatial correlation learning module, and a multi-scale gated convolution module. These components work together to capture complex temporal dependencies, dynamically model inter-series relationships as graphs, and fuse multi-scale spatial-temporal features, thereby enhancing MSCENet’s robustness and accuracy in detecting anomalies through dynamic feature interactions and multi-scale temporal relationships.
  \item Experiments conducted on real-world datasets show that MSCENet consistently exceeds the performance of existing deep learning models in the task of time series anomaly detection. Furthermore, MSCENet demonstrates exceptional generalization ability.
\end{itemize}

The paper is structured as follows: Section \ref{sec:related_works} provides a review of previous research, Section \ref{sec:methodology} describes our methodology in detail, Section \ref{sec:experiments} presents our experimental results and analysis, and Section \ref{sec:conclusion} offers a summary and conclusion of our findings.

\section{RELATED WORK}
\label{sec:related_works}
This section provides an overview of recent advances in time series anomaly detection and inter-series correlation learning.

\subsection{Detecting Anomalies in Multivariate Time Series}

Anomaly detection in time series has been extensively studied.Traditional statistical and machine learning approaches, such as ARIMA/VAR \cite{yu2016arima}, PCA \cite{LI201463}, SVM \cite{Manevitz2002}, wavelet analysis, and distance-based methods \cite{Lu2009,Wang2018}, often fall short in handling the high-dimensional, complex dependencies characteristic of modern multivariate time series.

Recent years have seen substantial progress in deep learning-based approaches, which address the limitations of traditional methods by better capturing non-linear patterns and complex dependencies across multiple variables. Deep One-Class Classification (Deep-SVDD) models, for example, encapsulate normal time series behavior within a hypersphere, with anomalies detected as data points lying outside this boundary \cite{ruff2018deep}. However, this approach assumes a consistent distribution of normal data, making it less effective in cases of concept drift. Encoder-Decoder Anomaly Detection (EncDecAD) employs Long Short-Term Memory (LSTM) networks to learn sequence features through reconstruction tasks, allowing for anomaly detection via reconstruction errors \cite{10090260}. While EncDecAD handles temporal dependencies well, it may struggle with capturing complex inter-variable relationships. CTAD adopts the one-class contrastive loss to address the issue that traditional contrastive loss misjudges temporally related samples, yet the synthesis of its negative data augmentation still requires optimization\cite{10325644}. In contrast, adversarial methods such as Unsupervised Anomaly Detection (USAD) \cite{Audibert2020USAD} enhances reconstruction quality, yet their focus on adversarial training introduces computational complexity, which can be a limitation for real-time applications. 

More recent advancements leverage stochastic and attention-based methods to capture long-term dependencies and subtle correlations across variables. OmniAnomaly \cite{su2019robust} uses stochastic representations to address the inherent complexity in multivariate time series, particularly for tasks requiring robust handling of temporal dependencies. Transformers, with their self-attention mechanism, offer further improvements by identifying long-term interactions in time series data \cite{chen2021autoformer}. Notably, TranAD \cite{TranAD2022} and the Anomaly Transformer \cite{xu2022anomaly} employ self-regulating anomaly attention mechanisms with a min-max strategy to minimize reconstruction errors in normal data while amplifying deviations in anomalous data, leading to strong performance gains. Despite their advancements, current methods are often limited by biases and suboptimal anomaly scores, particularly in scenarios affected by concept drift.

\subsection{Enhancing Inter-series Correlation Detection with GNNs}
Graph Neural Networks (GNNs) \cite{NIPS2016_04df4d43, kipf2017semisupervised, abu2019mixhop} have gained increasing prominence for modeling inter-series correlations, particularly in applications where capturing temporal dependencies is essential. These advancements illustrate GNNs' capability to model complex relationships within multivariate data, but they also highlight certain limitations in their current configurations. Most existing GNN architectures assume a predefined graph structure based on static inter-node relationships. While this structure is effective for domains where spatial relationships are stable and well-defined, it becomes problematic for broader multivariate forecasting tasks where establishing a universal and static graph structure is often infeasible. 

In fields such as finance, industrial monitoring, and environmental analysis, correlations between variables can vary significantly over time due to fluctuating conditions and context-dependent interactions, necessitating a more flexible graph structure to adapt to evolving inter-series relationships. To address these issues, recent research has explored adaptive, learnable graph structures \cite{Wu2019GraphWaveNet, WENG2023109670, wuConnecting2020} that seek to dynamically adjust inter-node relationships based on data-driven insights. Although these approaches represent a promising shift towards more flexible graph construction, they are often limited by the narrow range of graph structures they consider and by difficulties in adapting to temporal variations within inter-series correlations. Specifically, these adaptive GNNs frequently struggle to capture correlations that shift across different time scales, which can be crucial for accurately modeling complex and dynamic dependencies in multivariate time series data. 

\begin{figure*}[thb]
    \centering
    \includegraphics[width=0.98\textwidth]{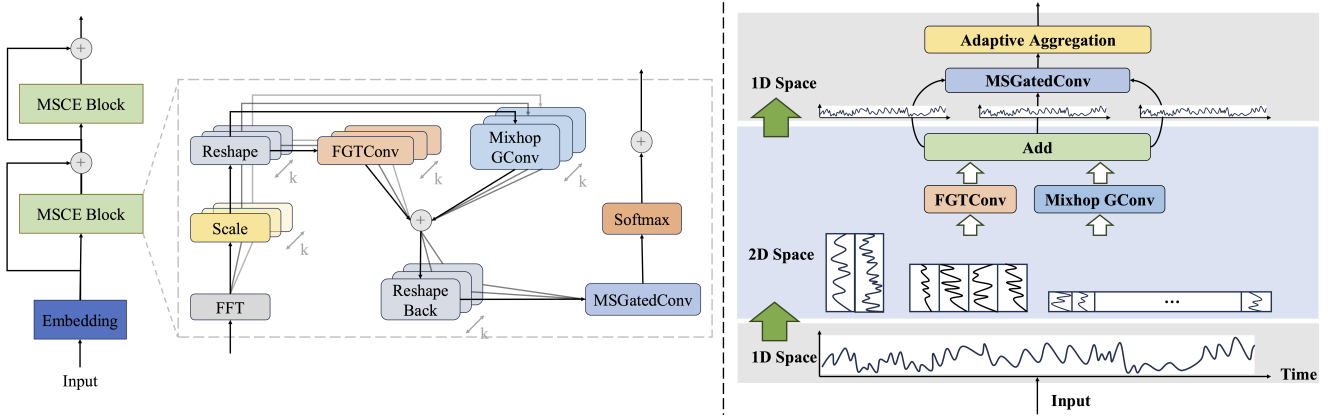}
    \caption{The overall architecture of MSCENet, illustrating the multi-scale temporal analysis, spatial correlation modeling via Mixhop graph convolutions, fine-grained temporal convolution, and multi-scale gated convolution mechanisms for comprehensive spatio-temporal anomaly detection.}
    \label{fig:architecture}
\end{figure*}
 
\section{METHODOLOGY}
\label{sec:methodology}
In this section, we propose the MSCENet framework and its detailed components, focusing on their role in detecting anomalous events in multivariate time series. The section begins by establishing a problem statement focused on labeling anomalies in a streaming environment, followed by an input embedding process that leverages both convolutional operations and residual connections for enhanced feature representation. As illustrated in Fig. \ref{fig:architecture}, to further enhance temporal analysis, we first perform multi-scale temporal analysis, identifying dominant time scales through Fourier transforms to enrich anomaly detection accuracy. Following this, our model introduces a fine-grained temporal convolution (FGTConv) module, which captures intricate temporal dependencies across multiple scales using dilated convolutions. Next, spatial correlations between series are modeled using graph structures, where Mixhop graph convolutions dynamically learn inter-series relationships, strengthening the model’s ability to capture complex spatial interactions. Finally, a multi-scale gated convolution mechanism fuses spatial and temporal features across diverse receptive fields, allowing the model to capture both local and global spatio-temporal dependencies effectively. This structured approach enables MSCENet to comprehensively analyze temporal and spatial patterns, significantly enriching anomaly detection accuracy. Anomalies are ultimately detected via reconstruction error, where deviations between reconstructed and original time series values signal potential anomalies. This hierarchical approach ensures our model effectively identifies and interprets anomalies in multivariate time series data, capturing both local and global patterns across different temporal scales and addressing the critical need to model complex, multi-scale inter-series dependencies that are often overlooked by traditional methods.

\subsection{Problem Statement}
\label{sec: problem_statement}
Before discussing the detailed aspects of the components previously described, the problem statement for this study is introduced. In a streaming environment, we analyze historical time series data $\mathcal{X}_T = \{x_1, x_2, \dots, x_t\}$, containing $N$ variables and observations with size $T$. For each time point $i$, we have $x_i \in \mathbb{R}^N$ and $\mathcal{X}_T \in \mathbb{R}^{N \times T}$, with $x_i = \{x_i^1, x_i^2, \dots, x_i^N\}$ denoting all observed variables. The aim is to label each point in $L = \{l_1, l_2, \dots, l_t\}$, where $l_i \in \{0, 1\}$, with a label of “1” for anomalies and “0” for normal data points.

\subsection{Embedding Inputs and Integrating Residual Connections}
\label{sec: embedding_input}
At a specific time step, a vector $x_i$ with size of $N$ is converted into an embedded representation $\hat{x}_i$, where the transformation is expressed as $x_i^N \to \hat{x}_i^{d_\text{model}}$. As a result, the overall embedding for the entire time series is represented as $\mathcal{X}_{T_\text{emb}} \in \mathbb{R}^{d_\text{model} \times T}$, with $d_\text{model}$ signifying the model's dimensionality and $T$ denoting the length of the time series. We refer to the standardized input structure presented in the Informer framework \cite{zhou2021informer} to effectively produce the embedding. In particular, $\mathcal{X}_{T_\text{emb}}$ is calculated with the following equation:
\begin{equation}
\label{deqn_ex1}
\mathcal{X}_{T_\text{emb}}=\alpha\mathrm{Conv}1\mathrm{D}(\mathcal{\hat{X}}_{T})+\mathbf{PE}+\sum_{p=1}^{P}\mathbf{SE}_{p}.
\end{equation}
At the beginning, we standardize the input $\mathcal{X}_{T}$ to result in $\mathcal{\hat{X}}_{T}$, because this normalization process is known to enhance stationarity \cite{liu2022non}. In the subsequent phase, $\mathcal{\hat{X}}_{T}$ is converted into a matrix of $d_\text{model}$ dimensions by applying 1-D convolution filters, which utilize 3 kernel widths and 1 stride length. The parameter $\alpha$ is utilized to align the scalar projection's magnitude with the embeddings at fine and coarse levels. The positional embedding for input $\mathcal{X}_T$ is denoted as $\mathbf{PE} \in \mathbb{R}^{d_\text{model} \times T}$, providing each time step with a unique position-based encoding, which aids in capturing sequential dependencies. Additionally, the global timestamp embedding $\mathbf{SE}_{p} \in \mathbb{R}^{d_\text{model} \times T}$ is introduced, which is learnable and constrained to a vocabulary of 60, as minutes are recognized as the smallest unit of specificity. Here, the index $p$ represents different periodic or seasonal patterns embedded in the data, and the model sums these embeddings across $P$ distinct periodic features. This summation of multiple timestamp embeddings $\sum_{p=1}^{P}\mathbf{SE}_{p}$ allows the model to capture various temporal characteristics, such as daily, weekly, and monthly patterns, enhancing its ability to model complex time dependencies effectively.

MSCENet utilizes a residual approach for its implementation \cite{he2016deep}. At the start, we establish $\mathcal{X}^0_T = \mathcal{X}_{T_\text{emb}}$, where $\mathcal{X}_{T_\text{emb}}$ denotes the raw input data enriched into features through the embedding matrix. In the $l$-th layer, the input $\mathcal{X}^{l-1}_T \in \mathbf{R}^{d_\text{model}\times T}$ is processed as shown below:
\begin{equation}
\mathcal{X}^l_T=\text{MSCEBlock}\left(\mathcal{X}^{l-1}_T\right)+\mathcal{X}^{l-1}_T,
\end{equation}
where MSCEBlock indicates the operations and calculations integral to the primary functions of the MSCENet layer.

\subsection{Multi-scale Temporal Analysis}
This study focuses on enhancing the accuracy of anomaly detection through the exploration of correlations among multiple time series at different temporal scales. The choice of suitable time scales is a vital component of this methodology. We argue that periodicity, a specific kind of time scale, is a significantly useful resource for selecting scales. Our interest in periodicity arises from its critical significance in analyzing time series data. By recognizing repeated patterns, periodicity offers crucial insights into the anomaly detection process. For instance, during the peak summer and winter months, a significant positive correlation is frequently found between electricity demand and temperature, which weakens during the milder periods of spring and fall. Selecting the appropriate time scale can significantly influence how this correlation is quantitatively assessed. Variations in periodicity, whether daily or monthly, may lead to different values for the correlation coefficients.

To identify the optimal time scale for analyzing time series anomalies, we consider this step crucial in our detection approach. This is accomplished through the use of the Fast Fourier Transform ($\mathrm{FFT}$), a method influenced by the research presented in TimesNet \cite{wu2023timesnet}:
\begin{equation}
\begin{aligned}
\mathbf{F}=\mathrm{Avg}\left(\mathrm{Amp}\left(\mathrm{FFT}(\mathcal{X}_{T}^{l-1})\right)\right), \\ f_{1},\cdots,f_{k}=\underset{f_{*}\in\{1,\cdots,\frac{T}{2}\}}{\operatorname*{argTopk}}(\mathbf{F}),s_{i}=\frac{T}{f_{i}},
\end{aligned}
\end{equation}
where $\mathrm{Amp}(\cdot)$ calculates the magnitude or strength of each resulting frequency component, representing their relative contributions. The vector $\mathbf{F}\in\mathbb{R}^{L}$ captures the amplitude values for each frequency, which are averaged across the $d_\text{model}$ dimensions using the $\mathrm{Avg}(\cdot)$ function to obtain a stable representation of the dominant frequency characteristics. In this context, $f_{1},\cdots,f_{k}$ are the indices of the top $k$ frequencies with the highest amplitudes in $\mathbf{F}$, determined by the $\operatorname*{argTopk}$ function, and each corresponding scale $s_i$ is calculated as $s_{i}=\frac{T}{f_{i}}$, providing interpretable periodic intervals relevant to the anomaly detection task.

This study emphasizes the importance of recognizing that temporally varying inputs may reveal identifiable periodic behaviors, which allows our model to detect changing scales. We suggest that the critical correlations within this evolving periodicity are stable over time. Such a viewpoint calls for an examination of the dynamic attributes found in the inter-series and intra-series correlations articulated by our model.

By utilizing the specified time scales ${s_{1},\ldots,s_{k}}$, we generate diverse representations for different temporal scales through the transformation of the input data into 3D tensors, as demonstrated by the subsequent equation:
\begin{equation}
\mathcal{X}^{i}_\mathrm{2D}=\mathrm{Reshape}_{s_i,f_i}(\mathrm{Padding}(\mathcal{X}_{T}^{l-1})),\quad i\in\{1,\ldots,k\},
\end{equation}
in this context, employing $\mathrm{Padding}(\cdot)$ allows the time series to be extended with zeros through the temporal axis, which ensures it aligns with $\mathrm{Reshape}_{s_{i},f_{i}}(\cdot)$. For the time scale $i$, $\mathcal{X}^{i}_\mathrm{2D} \in \mathbb{R}^{d_\text{model}\times s_{i}\times f_{i}}$ denotes the restructured $\mathrm{2D}$ tensor for the $i$-th scale.

\begin{figure}[thb]
    \centering
    \includegraphics[width=0.40\textwidth]{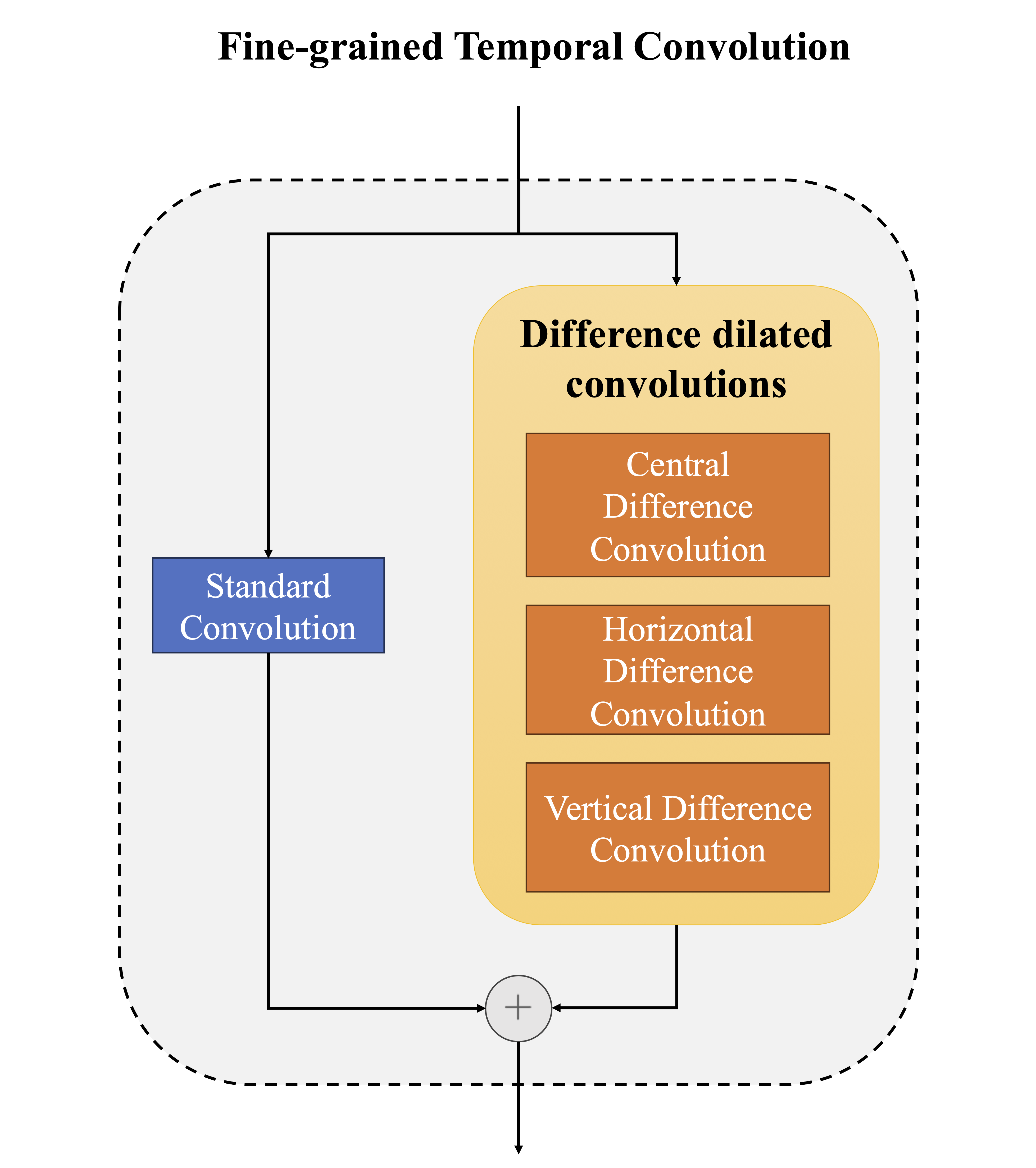}
    \caption{Architecture of the fine-grained temporal convolution (FGTConv) module, which consists of three parallel dilated convolutions and one standard convolution. The module employs central difference convolution (CDC), horizontal difference convolution (HDC), and vertical difference convolution (VDC) alongside the standard convolution, enabling multi-scale feature extraction by capturing both intensity and gradient information across temporal scales.}
    \label{fig:fine-grained_temporal_convolution}
\end{figure}

\subsection{Fine-grained Temporal Convolution}
\label{sec:fine-grained_temporal_convolution}
Recently, there has been growing interest in the use of image processing techniques to analyze time series data \cite{Li2023,10496248}. The key idea of this approach is to transform standard time series patterns into visual images, which can then be analyzed to reveal significant spatial features. This transformation simplifies the complexities associated with time series data and makes it possible to apply advanced image analysis techniques, including convolutional neural networks (CNNs) and other deep learning strategies \cite{wu2023timesnet, liu2024multivariate}. The process begins with normalizing the time series data to ensure that variations due to scale or noise do not obscure the underlying patterns. Transforming time series data into images allows for the extraction of strong spatial features that are not immediately apparent in the raw data. This process improves predictive modeling, anomaly detection, and classification, providing researchers and practitioners with clear insights into data dynamics.

Many previous methods rely on standard convolution layers for feature extraction and learning, exploring a vast solution space without constraints. These approaches often start with random initial conditions, which limit their modeling capabilities and expressiveness \cite{Wu2021, 10411857}. In contrast to these traditional techniques that directly obtain spatio-temporal features from raw data, our method adopts a different strategy. By addressing the limitations of standard convolutional networks, we introduce a fine-grained temporal convolution (FGTConv) specifically designed to capture intricate temporal relationships. This innovation enables a more effective balance between local and global characteristics over extended timeframes, enhancing the overall effectiveness of modeling for spatio-temporal data. 

Our architecture, as illustrated in Fig. \ref{fig:fine-grained_temporal_convolution}, comprises four convolutional layers, which include three dilated convolutions and one standard convolution, all operating concurrently to extract features. Atrous convolution, commonly known as dilated convolution, is a sophisticated technique that enables a model to expand its receptive field without increasing the number of parameters or the computational burden \cite{Yu2020, 8561240}. By introducing "gaps" (or zeros) between the elements of the kernel, dilated convolutions effectively capture a broader contextual representation from the input data, allowing the network to analyze a more extensive segment of the input signal. This methodology is particularly advantageous in time series analysis, where understanding dependencies across varying temporal scales is essential. The flexibility of adjustable dilation rates empowers the model to learn features at multiple scales, facilitating the detection of patterns that standard convolutions may overlook. Consequently, dilated convolutions enhance the model's ability to identify both local and global temporal dependencies, thereby leading to more effective feature extraction \cite{Wang2023, YANG2023106151}. In the formulation of our FGTConv, we apply several types of difference convolutions, including central difference convolution (CDC), horizontal difference convolution (HDC), and vertical difference convolution (VDC), to incorporate traditional local descriptors. CDC enhances the model's capacity by rearranging learned kernel weights to save computational cost and memory consumption \cite{Su2021}. HDC computes horizontal gradients by evaluating differences among selected pixel pairs, and the convolution kernel weights are rearranged after training for direct application to input features. Both HDC and VDC are designed to enhance gradient information, thus improving the model's representational and generalization capabilities. The standard convolution is used to extract intensity-level data, while the difference convolutions emphasize gradient-level insights. Let $\mathcal{X}^{l, i}_\mathrm{2D}$ the represent the input feature map,  the formulation can be expressed as follows:
\begin{equation}
\begin{gathered}
\mathcal{X}^{i}_\mathrm{FGTConv} = \mathrm{FGTConv
}(\mathcal{X}^{i}_\mathrm{2D})=\mathcal{X}^{i}_\mathrm{2D}*(\sum_{j=1}^4{w_j} {K_i}) \\
=\mathcal{X}^{i}_\mathrm{2D}*K_\text{combined}, 
\end{gathered}
\end{equation}
where $\mathcal{X}^{i}_\mathrm{FGTConv}$ denotes the output of the FGTConv operation, $\mathrm{FGTConv(\cdot)}$ represents our proposed FGTConv function. The parameters $w_j$ are the learnable weights corresponding to each convolution type. $K_j$ for $j = 1, \cdots, 4$ signifies the kernels associated with standard convolution, CDC, HDC, and VDC, respectively. Notably, all convolutional layers utilize kernels with a size of $3 \times 3$ and a stride of $1$, ensuring consistent spatial resolution across feature maps. The symbol $*$ indicates the convolution operation, while $K_\text{combined}$ represents the combined kernel that integrates the outputs of the parallel convolutions.

\subsection{Spatial Correlation Learning through Graph Structures}
\label{sec: spatial_correlation_learning}
\begin{figure}[thb]
    \centering
    \includegraphics[width=0.48\textwidth]{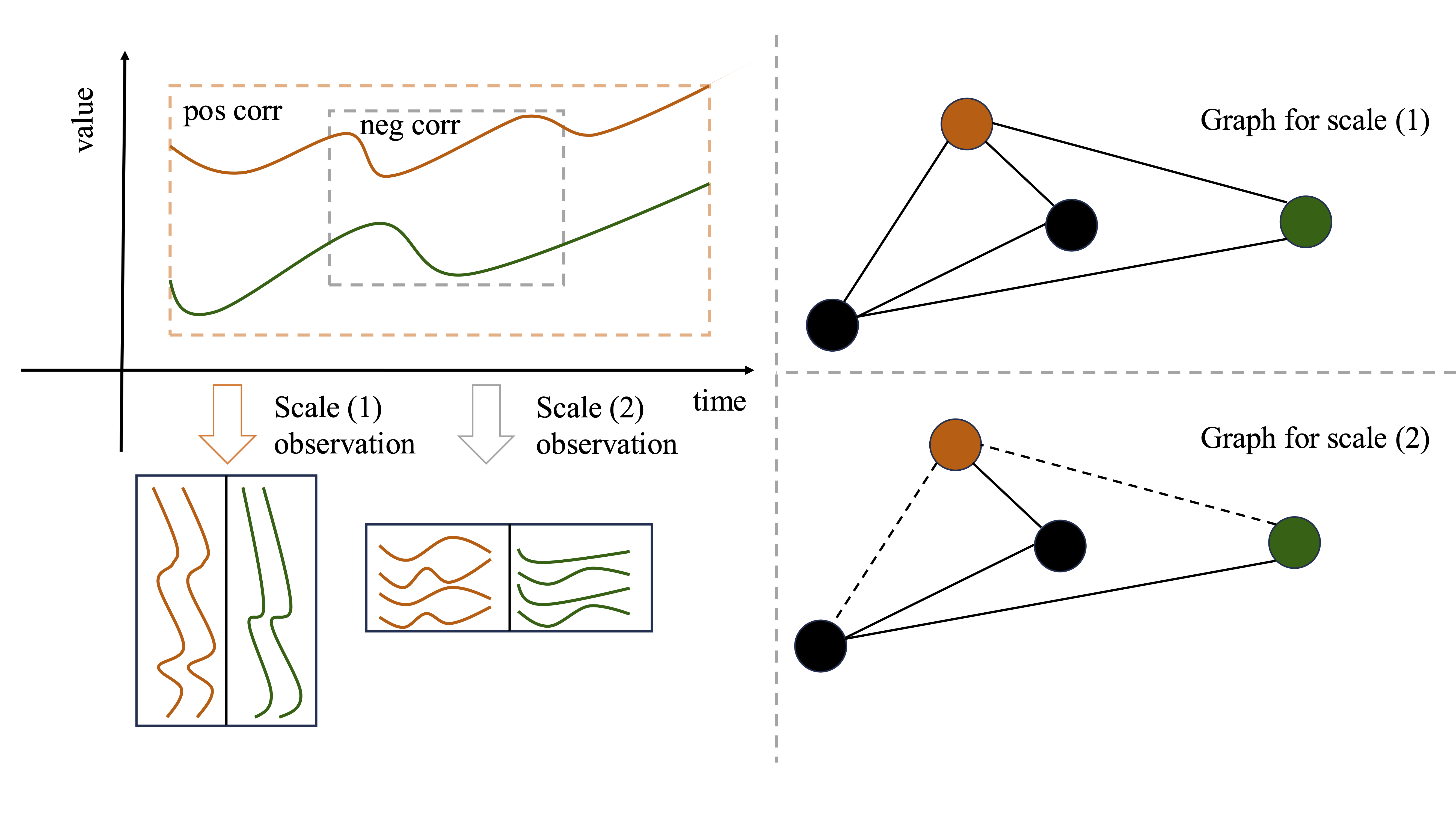}
    \caption{Analyzing the orange and green time series over an extended time series (1) reveals a positive correlation between them. However, on a shorter time scale (2), a negative correlation is observed between these series. This results in the formation of two distinct graph structures, each representing the different time scales.}
    \label{fig:spatial_correlation}
\end{figure}

Time series analysis often neglects the diverse inter-series relationships across multiple time scales, which current deep learning models fail to capture effectively. This oversight is critical for anomaly detection. Spatial correlations among physiological signals, such as heart rate, blood pressure, and respiratory rate, can vary significantly, strengthening during emergencies to indicate heightened alertness and weakening under stable conditions. Similarly, in industrial systems, sensor readings reveal spatial dependencies among sensors that intensify during anomalies and diminish during normal operations. Fig. \ref{fig:spatial_correlation} illustrates that at one time scale, two time series may show a positive spatial relationship, while at a more granular scale, a negative correlation might emerge. These variations in spatial correlations across different time scales serve as crucial indicators of anomalies, providing valuable insights for detection. To effectively model these spatial correlations, we employ a graph-based approach that constructs distinct graph representations of inter-series relationships across time scales. This methodology enhances our ability to detect anomalies, offering a more nuanced analysis than traditional models. By learning spatial correlations through graph structures, we gain a deeper understanding of the intricate dynamics among time series.

Time series data frequently exhibit intricate dependencies due to varying intra-series and inter-series correlations, which have been the focus of various studies \cite{Cai2024, 10589693}. To account for the influence of one instance's state on others, we propose a module that utilizes the Mixhop graph convolution method. This approach is designed to effectively capture the intricate inter-series relationships within time series data, thereby enhancing the accuracy of its representation.

To accomplish this objective, we initially revert the tensor linked to the $i$-th scale into a tensor encompassing $N$ variables.  The operation is implemented by means of a linear transformation, specified as follows: 
\begin{equation}
\mathrm{H}^i=\mathbf{P}^i \mathcal{X}^{i}_\mathrm{2D},
\end{equation}
where $\mathcal{H}^i \in \mathbb{R}^{N \times s_{i} \times f_{i}}$ represents the output tensor, while $\mathbf{P}^i$ is a learnable weight matrix specifically designed for the $i$-th scale tensor. This transformation allows for effective adaptation of the input data to the desired representation. As part of the graph learning process, we introduce two trainable parameter matrices $W_i^1$ and $W_i^2 \in \mathbb{R}^{N \times h}$, which are used to compute an adaptive adjacency matrix as follows: 
\begin{equation}
M^i = \text{SoftMax}(\text{ReLU}(W_i^1 (W_i^2)^T)),
\end{equation}
where $M^i$ represents the adjacency matrix for the $i$-th scale.  To ensure a balanced representation of inter-series relationships, the $\text{SoftMax}(\cdot)$ function normalizes the weights between nodes. Inter-series correlations are then detected using the Mixhop graph convolution technique, employing the adjacency matrix $M^i$ as a foundational component. We present the definition of graph convolution in the following:
\begin{equation}
\mathrm{H}_\text{out}^i = \sigma \left( \bigg\|_{j \in \beta} (M^i)^j \mathrm{H}^i \right),
\end{equation}
where $\mathrm{H}_\text{out}^i$ denotes the output resulting from combination at scale $i$, the hyper-parameter $\beta$ comprises a collection of integer adjacency powers, $(M^i)^j$ represents the learned adjacency matrix $M^i$ raised to the $j$-th power, \(\parallel\) signifies a column-wise concatenation, which concatenates intermediate variables produced within each iteration, and $\sigma$ serving as the activation function. Following this procedure, we transform $\mathrm{H}_\text{out}^i$ back into a 3D tensor $\mathcal{X}_{\mathrm{GCN}}^{i} \in \mathbb{R}^{d_\text{model} \times s_i \times f_i}$ leveraging a multi-layer perceptron.

\subsection{Multi-scale Gated Convolution for Spatial-Temporal Feature Extraction}
\begin{figure}[thb]
    \centering
    \includegraphics[width=0.48\textwidth]{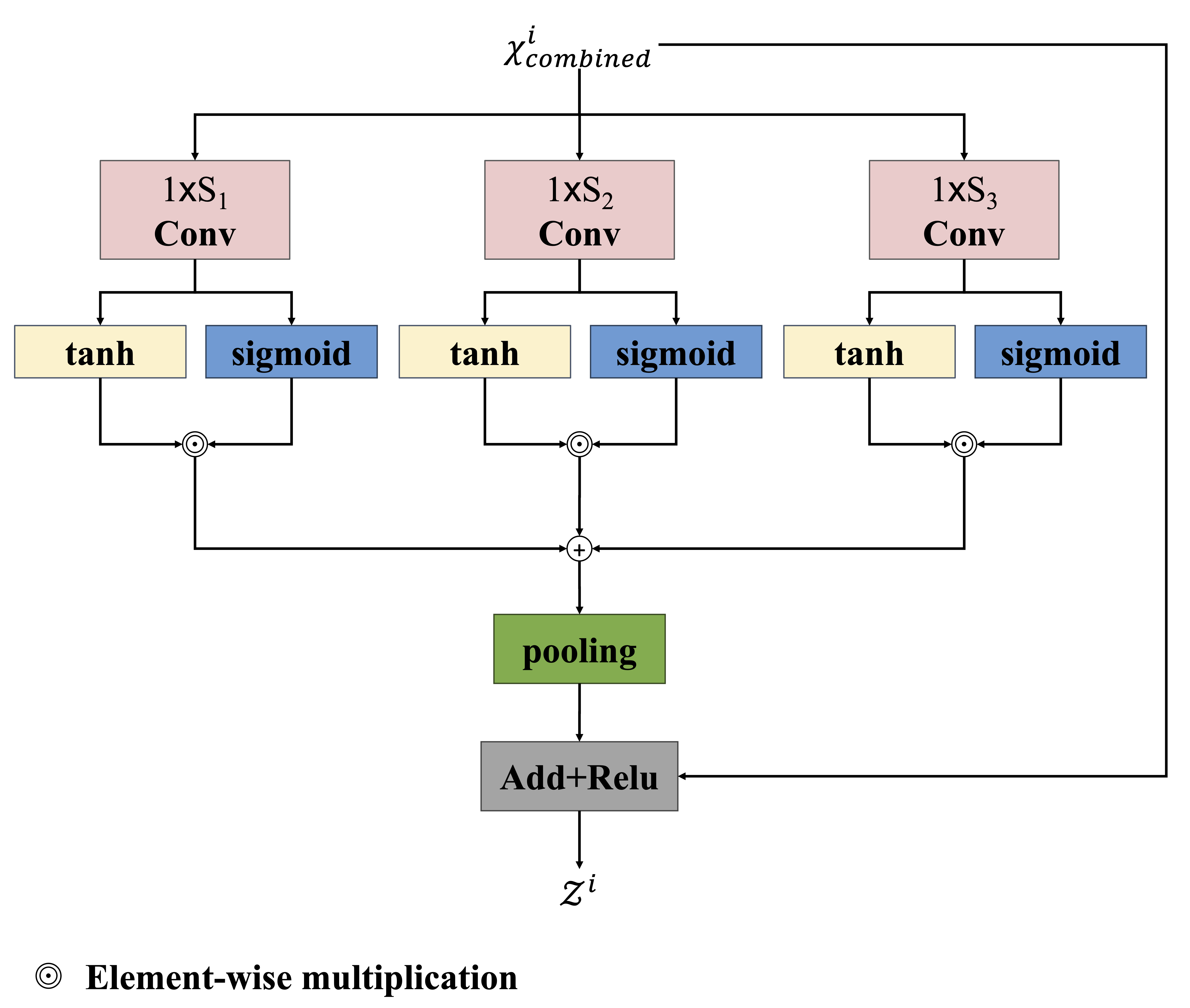}
    \caption{Architecture of the multi-scale gated convolution module, incorporating GTUs with varying receptive fields for enhanced spatial-temporal feature extraction.}
    \label{fig:M_GTU}
\end{figure}

In contrast to the TSSRGCN approach \cite{9338393}, which employs a cycle-based dilated deformable convolution to effectively capture both long-term and short-term temporal dynamics in traffic data, we present a novel approach utilizing multi-scale gated convolution to effectively fuse spatial and temporal features in time series analysis. In Fig. \ref{fig:M_GTU}, the specific architecture of the module is presented, composed of three Gated Tanh Unit (GTU) \cite{dauphin2017language} modules that have differing receptive fields \cite{lan2022dstagnn}. Before extracting multi-scale features, we first combine the outputs from Section. \ref{sec:fine-grained_temporal_convolution} and Section. \ref{sec: spatial_correlation_learning}. This integration can be mathematically represented as:
\begin{equation} 
\mathcal{X}^{i}_\text{combined} = \mathcal{X}^{i}_\mathrm{FGTConv} + \mathcal{X}_{\mathrm{GCN}}^{i}, 
\end{equation}
the summation effectively merges the spatial information captured through Mixhop graph convolutions with the temporal features extracted by FGTConv.

Subsequently, we apply a series of 1D causal convolutions with different kernel sizes $S_1$, $S_2$, $S_3$ on $\mathcal{X}^{i}_\text{combined}$ to extract multi-scale features from the time series data. This methodology enables the model to capture dependencies at various temporal scales, thereby offering an integrated portrayal of the temporal dynamics. Simultaneously, the channel dimension is projected to $2d_\text{model}$ to enhance feature capacity. The resulting feature maps are computed as follows:
\begin{equation} 
\begin{aligned} 
\mathcal{X}_{S_j}^i = \mathrm{Conv}_{1\times S_j}(\mathcal{X}^{i}_\text{combined}), \quad j \in \{1, 2, 3\} \ 
\end{aligned} 
\end{equation}
where $\mathcal{X}_{S_j}^i \in \mathbb{R}^{2d_\text{model} \times s_i \times (f_i - S_j + 1)}$, and $S_j$ corresponds to the kernel sizes of $1\times3$, $1\times5$, and $1\times7$ for $j \in \{1, 2, 3\}$ respectively, in our implementation.  By leveraging multiple receptive fields through these different kernel sizes, the model gains the flexibility to identify patterns that may be otherwise overlooked with a single fixed-scale convolution. This hierarchical, multi-scale approach enhances the model's representation of temporal dynamics, supporting more robust and nuanced time series analysis. To regulate the flow of information and introduce nonlinearity, we implement a gate mechanism through the use of GTUs. Specifically, we partition the input tensor along the channel dimension into two equal parts, allowing for selective modulation of information as below: 
\begin{equation}
\mathcal{E}, \mathcal{F}=\mathrm{Split}(\mathcal{X}_{S_j}^i),
\end{equation}
where $\mathcal{E}$, $\mathcal{F} \in \mathbb{R}^{d_\text{model}\times s_i \times(f_i-S_j+1)}$ represent the two halves of the feature map $\mathcal{X}_{S_j}$ after it has been split along the channel dimension. We then define the output of the GTU using a gating mechanism as follows:
\begin{equation}
\mathcal{\hat{X}}_{S_j}^i=\tanh(\mathcal{E})\odot\sigma(\mathcal{F}),
\end{equation}
where $\mathcal{\hat{X}}_{S_j}^i \in \mathbb{R}^{d_\text{model}\times s_i \times(f_i-S_j+1)}$ contains the gated multi-scale features, while $\tanh(\cdot)$ is the activation function applied to $\mathcal{E}$ and $\sigma(\cdot)$ generates gating weights from $\mathcal{F}$. The symbol $\odot$ denotes element-wise multiplication , which combines $\tanh(\mathcal{E})$ and $\sigma(\mathcal{F})$ to selectively allow information to pass based on the gating values. After obtaining the gated multi-scale features $\mathcal{\hat{X}}_{S_j}$ for each receptive field size, we concatenate these feature maps along the temporal dimension to integrate information across different scales. This operation can be expressed as:
\begin{equation} 
\mathcal{H}^i = \mathrm{CONCAT}(\mathcal{\hat{X}}_{S_1}^i, \mathcal{\hat{X}}_{S_2}^i, \mathcal{\hat{X}}_{S_3}^i), 
\end{equation}
where $\mathcal{H}^i \in \mathbb{R}^{d_\text{model}\times s_i \times(3f_i-{S_1}-{S_2}-{S_3}+3)}$ contains the combined multi-scale features across all three receptive fields. This concatenation along the temporal dimension provides a comprehensive representation that captures both short-term and long-term dependencies within the time series data. Following the concatenation, we apply a max pooling layer with a kernel size $W=2$ along the temporal dimension to further reduce the dimensionality and focus on the most relevant temporal information:
\begin{equation} 
\mathcal{H}_\text{pooled}^i = \mathrm{Pooling}(\mathcal{H}^i), 
\end{equation}
where $\mathcal{H}_{\text{pooled}}^i\in\mathbb{R}^{d_\text{model}\times s_i \times\frac{(3T-(S_1+S_2+S_3-3))}W}$. This pooling operation helps compress the temporal dimension while retaining essential patterns from the concatenated features. To maintain consistency with the original time dimension, a linear transformation could be applied to $\mathcal{H}_\text{pooled}^i$ to directly project it back to the original time dimension $T$. This step ensures compatibility with downstream layers and helps retain the temporal resolution needed for effective analysis. 

Finally, we add a residual connection by combining $\mathcal{H}_\text{pooled}^i$ with the original input $\mathcal{X}^{i}_\text{combined}$, and a ReLU activation is applied to obtain the output from the multi-scale gated convolution module:
\begin{equation}
\mathcal{Z}^i=\mathrm{ReLU}(\mathcal{H}_\text{pooled}^i+\mathcal{X}^{i}_\text{combined}),
\end{equation}
where $\mathcal{Z}^i \in \mathbb{R}^{d_\text{model}\times s_i \times f_i}$. To move to the next layer, we aggregate the $k$ scale tensors, $\mathcal{Z}^1, \cdots, \mathcal{Z}^k$. First, each tensor is reshaped into a 2-way matrix $\hat{\mathcal{Z}}^i \in \mathbb{R}^{d_\text{model}\times T}$, then the scales are combined according to their amplitudes:
\begin{equation}
\begin{aligned}
\hat{a}_{1},\cdots,\hat{a}_{k}=& \mathrm{SoftMax}(\mathbf{F}_{f_{1}},\cdots,\mathbf{F}_{f_{k}}), \\
\mathcal{\hat{Z}}=& \sum_{i=1}^k\hat{a}_i\mathcal{\hat{Z}}^i.
\end{aligned}
\end{equation}
For each scale $i$, the FFT is applied to compute the amplitudes $\mathbf{F}_{f{1}},\cdots,\mathbf{F}_{f{k}}$, and the $\text{SoftMax}(\cdot)$ function is then used to compute the final amplitudes $\hat{a}_{1},\cdots,\hat{a}_{k}$.

\subsection{Anomaly Detection via Reconstruction Error}
By measuring the reconstruction error, which compares the original data with the model's output, we can detect anomalies in time series data. Larger errors highlight abnormal behaviors, enhancing the detection process. To begin the process, a feed-forward network is applied to each reshaped tensor to map it into the reconstruction space $\hat{\mathcal{Y}}_t \in \mathbb{R}^N$, as follows:
\begin{equation}
\hat{\mathcal{Y}}_t=\text{FeedForward}(\mathcal{\hat{Z}}).
\end{equation}
During training, the sensor embeddings and other neural network weights are updated based on the mean squared error between $\mathcal{Y}_t$ and $\hat{\mathcal{Y}}_t$, without the computation of anomaly scores. In the testing phase, the L2 loss between the predicted is computed.
\begin{equation}
\mathrm{AnomalyScore}_t= \frac1N\sum_{n=1}^N\left(\left(\|\hat{\mathcal{Y}}_t-\mathcal{Y}_t\|_2\right)\right).
\end{equation}
Ultimately, when the anomaly score exceeds the preset threshold, we classify the time step $t$ as an anomaly. The threshold is determined by the mean anomaly score on the validation set in our approach. The proportion of training data to validation data reflects the degree to which anomalies are spread throughout the dataset.

\section{Experiments}
\label{sec:experiments}
A detailed analysis of MSCENet is carried out using three benchmark datasets, focusing on its application in three practical use cases. 

\subsection{Datasets}
The data for experimentation is drawn from three unique sources, each with specific dimensional characteristics and properties.

\begin{itemize}
  \item SMD (Server Machine Dataset) \cite{su2019robust}: Collected over a period of five weeks, the resource utilization of different machines in a large-scale computing cluster at an online company is recorded in this dataset.
  \item PSM (Pooled Server Metrics) \cite{abdulaal2021practical}: This dataset, with 26 different dimensions, is sourced from various eBay application server nodes within the internal network.
  \item SWaT (Secure Water Treatment) \cite{mathur2016swat}: This dataset originates from 51 sensors situated in a vital facility system, which operates continuously without any interruptions.
\end{itemize}

\input{table/datasets_detail}

\subsection{Metric}
The performance of time series anomaly detection is commonly evaluated using three key metrics: precision, recall, and F1-score (F1). For detecting anomalies, we follow the established point adjustment method, marking the entire segment as anomalous when any data point in it is flagged. Consequently, our focus is on detecting anomalies in time series without supervision. Unsupervised anomaly detection in time series encounters substantial challenges in practical scenarios. The system needs to extract meaningful representations from intricate temporal patterns using unsupervised methods and formulate a unique criterion that can identify infrequent anomalies amidst numerous normal data points.

\subsection{Baselines}

\begin{itemize}
    \item Isolation Forest \cite{4781136}: In unsupervised anomaly detection, Isolation Forest isolates anomalies by building random decision trees, where anomalies are identified due to their shorter path lengths. For time series data, it uses sliding windows to compute anomaly scores, making it effective for high-dimensional datasets, though sensitive to hyper parameters.
    \item Deep-SVDD \cite{ruff2018deep}: Deep-SVDD earns latent representations through a deep neural network and applies a spherical boundary for anomaly detection. However, Deep-SVDD optimizes the latent representations and the anomaly detection boundary separately, which may limit the alignment between the representations and the anomaly detection task.
    \item OmniAnomaly \cite{su2019robust}: OmniAnomaly is a probabilistic recurrent neural network that captures standard patterns in multivariate time series by creating strong representations, utilizing methods such as stochastic variable connections and planar normalizing flow. Anomalies are identified based on reconstruction probabilities, and OmniAnomaly also provides interpretability by analyzing the reconstruction probabilities of each univariate component. 
    \item GDN \cite{Deng2021}: GDN learns feature dependencies via graph structure learning and uses GAT to derive temporal features, functioning as a traditional prediction-oriented anomaly detection technique.
    \item InterFusion \cite{li2021multivariate}: By utilizing a hierarchical Variational AutoEncoder with two stochastic latent variables, InterFusion is an unsupervised method that models inter-metric and temporal dependencies in multivariate time series, capturing low-dimensional embeddings for each of these dependencies. It also employs an MCMC-based method for anomaly interpretation by localizing the most anomalous metrics.
    \item TranAD \cite{TranAD2022}: TranAD combines Transformer architecture for efficient sequence encoding, adversarial training for minimizing reconstruction errors, and model-agnostic meta-learning to enhance the speed of inference, offering an integrated solution to anomaly detection.
    \item TimesNet \cite{wu2023timesnet}: By using a 2D mapping of 1D time series, TimesNet captures detailed time-based changes over several periods. By embedding intraperiod and interperiod variations into 2D tensors, TimesNet leverages 2D kernels for effective temporal modeling, making it suitable for forecasting, classification, and anomaly detection.
\end{itemize}

\input{table/baseline_comparison}

\subsection{Experiments Setup}
According to the pre-processing strategy in Anomaly Transformer \cite{xu2022anomaly}, using a sliding window, we divide the dataset into consecutive, non-overlapping sections. We set a sliding window size of 100 to capture sequential patterns effectively. For all datasets, embedding vectors of length 64, a batch size of 128, and hidden layers with 64 neurons are employed to ensure a balanced model complexity. The Adam optimizer with a learning rate of $1 \times 10^{-3}$ is used to ensure stable convergence, while training is limited to 20 epochs, and early stopping is implemented with a patience of 5 epochs. For enhanced accuracy, the top-K is set to 5, the Mixhop order is set to 2, and the output layers are configured to 2. Baseline data and code are sourced from relevant papers and official repositories.

Experiments are performed using an Ubuntu 20.04 operating system, powered by a 15 vCPU Intel(R) Xeon(R) Platinum 8474C @ 2.1 GHz CPU, 80 GB RAM, and a GeForce RTX 4090D (24GB) graphics card. 

\subsection{Experimental Results}
Table \ref{tab:baseline_comparison} summarizes the experimental results, providing a comparison of the performance of different anomaly detection methods on the SMD, PSM, and SWaT datasets. Our proposed method consistently outperforms existing approaches across all metrics, showcasing the effectiveness of the multi-scale correlation enhanced framework. To further explain these findings, we offer a detailed comparison and explore the specific contributions of our model, providing insights into each dataset and contrasting it with baseline methods.

For the SMD dataset, traditional methods like Isolation Forest struggle with these dependencies, achieving a relatively low F1-score of 53.64\%, with a precision of 42.31\% and recall of 73.29\%. This disparity indicates that Isolation Forest struggles to balance between detecting true anomalies and minimizing false positives, reflecting its limited capability for handling multivariate time series data. More advanced models like Deep-SVDD and GDN achieve improved performance, with F1-scores of 79.10\% and 83.42\%, respectively. Deep-SVDD, with a precision of 78.54\% and recall of 79.67\%, demonstrates a balanced approach but still lacks depth in capturing complex inter-series relationships. Although GDN achieves a recall rate of 99.74\%, its precision of 71.70\% indicates that it suffers from a high number of false positives, likely because of its limited ability to capture the complex interdependencies between variables. In contrast, models like OmniAnomaly, InterFusion, and TimesNet benefit from temporal modeling capabilities, achieving F1-scores above 84\%. Specifically, OmniAnomaly reaches a precision of 83.68\% and recall of 86.82\%, showing a balanced performance. However, these models lack multi-scale analysis and dynamic spatial learning, limiting their ability to adapt to varying inter-series correlations over time. MSCENet, with its innovative integration of multi-scale analysis and graph-based spatial learning, efficiently captures both transient and sustained dependencies. This design enables MSCENet to achieve a significantly higher F1-score of 86.68\%, with a precision of 88.24\% and recall of 85.71\%. 

For the PSM dataset, traditional approaches like Isolation Forest demonstrate limited capability, with an F1-score of 83.48\%, a precision of 76.09\%, and a high recall of 92.45\%. While Isolation Forest manages to detect a large portion of anomalies, its relatively low precision suggests that it generates numerous false positives, limiting its reliability. Advanced models such as Deep-SVDD and TranAD exhibit differing recall rates, with Deep-SVDD achieving a relatively strong recall of 86.49\%, indicating its capacity to identify anomalies, while TranAD’s more moderate recall of 76.74\% suggests it may miss some anomalies in the dataset. Although Deep-SVDD and TranAD achieve F1-scores of 90.73\% and 85.17\%, respectively, along with corresponding precisions of 95.41\% and 95.67\%, their overall performance remains limited in handling the complex temporal structures of the dataset, which may lead to either missed anomalies or false positives in certain cases. TimesNet significantly outperforms InterFusion on the PSM dataset, with an F1-score of 94.45\% compared to InterFusion's 83.52\%. TimesNet achieves a high precision of 98.56\% and recall of 90.66\%, demonstrating its strength in temporal modeling, though it lacks flexibility for dynamic inter-series relationships. InterFusion, with a precision of 83.61\% and recall of 83.45\%, offers a more balanced but less optimal performance due to its lack of a multi-scale or graph-based approach. MSCENet, by contrast, achieves a superior F1-score of 97.29\%, with a precision of 98.44\% and recall of 96.17\%. This strong balance between precision and recall underscores MSCENet's ability to leverage multi-scale temporal feature extraction and dynamic spatial learning, allowing it to adapt to evolving dependencies across time and features.

For the SWaT dataset, advanced methods such as GDN and OmniAnomaly achieve reasonable F1-scores of 81.01\% and 82.83\%, respectively. Notably, GDN has a high precision of 96.97\%, indicating effectiveness in minimizing false positives. However, its low recall of 69.57\% suggests that it misses a significant number of true anomalies, reflecting its limitations in capturing all anomalous patterns within the dataset. OmniAnomaly, on the other hand, achieves a more balanced performance with a precision of 81.42\% and a recall of 84.30\%, but its F1-score remains lower due to difficulties in adapting to SWaT’s complex dependencies and varying temporal scales. More advanced models like InterFusion, TranAD, and TimesNet demonstrate an improved balance between precision and recall. InterFusion achieves an F1-score of 83.01\%, with a precision of 80.59\% and recall of 85.58\%, while TranAD attains an F1-score of 88.28\%, with precision and recall of 87.99\% and 88.58\%, respectively. TimesNet, with its strong temporal modeling capabilities, achieves a high F1-score of 88.90\%, precision of 86.32\%, and recall of 91.64\%. While these models perform well overall, they still lack the robustness required to consistently handle dynamic inter-series relationships across diverse time scales, leading to either missed anomalies or false positives. In contrast, MSCENet achieves a superior F1-score of 90.09\%, with a precision of 97.53\% and recall of 83.70\%, significantly outperforming existing methods. This careful balance of high precision and recall demonstrates MSCENet's remarkable ability to accurately detect anomalies, significantly reducing both false positives and instances of missed detections. Its multi-scale gated convolution mechanism allows it to effectively adapt to different temporal scales, capturing both short-term and long-term dependencies, making it highly effective for detecting anomalies.

\begin{figure}[thb] \centering
    \includegraphics[width=0.40\textwidth]{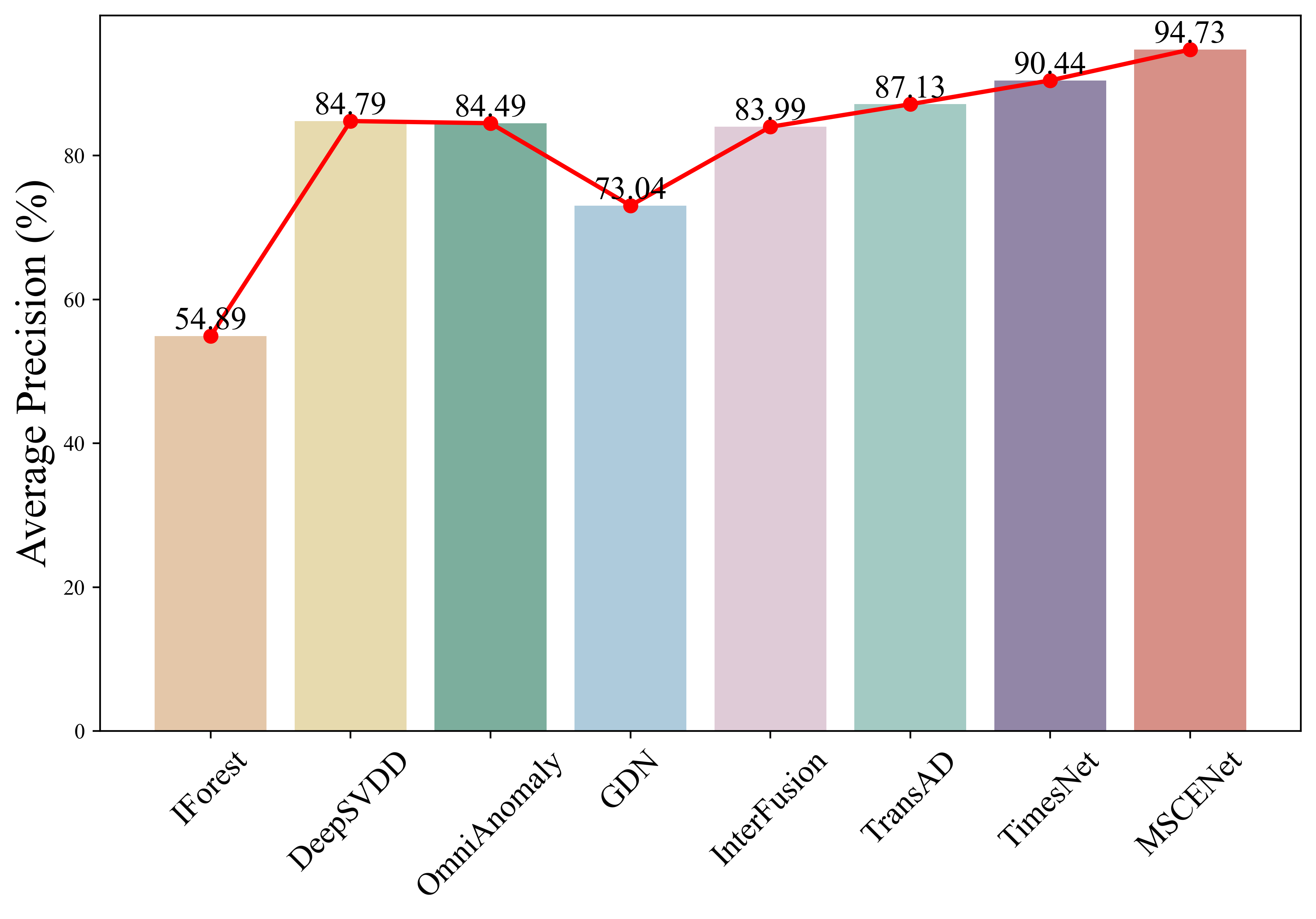}
    \\
    \makebox[0.45\textwidth]{\small (a) Average Precision Comparison}
    \\
    \includegraphics[width=0.40\textwidth]{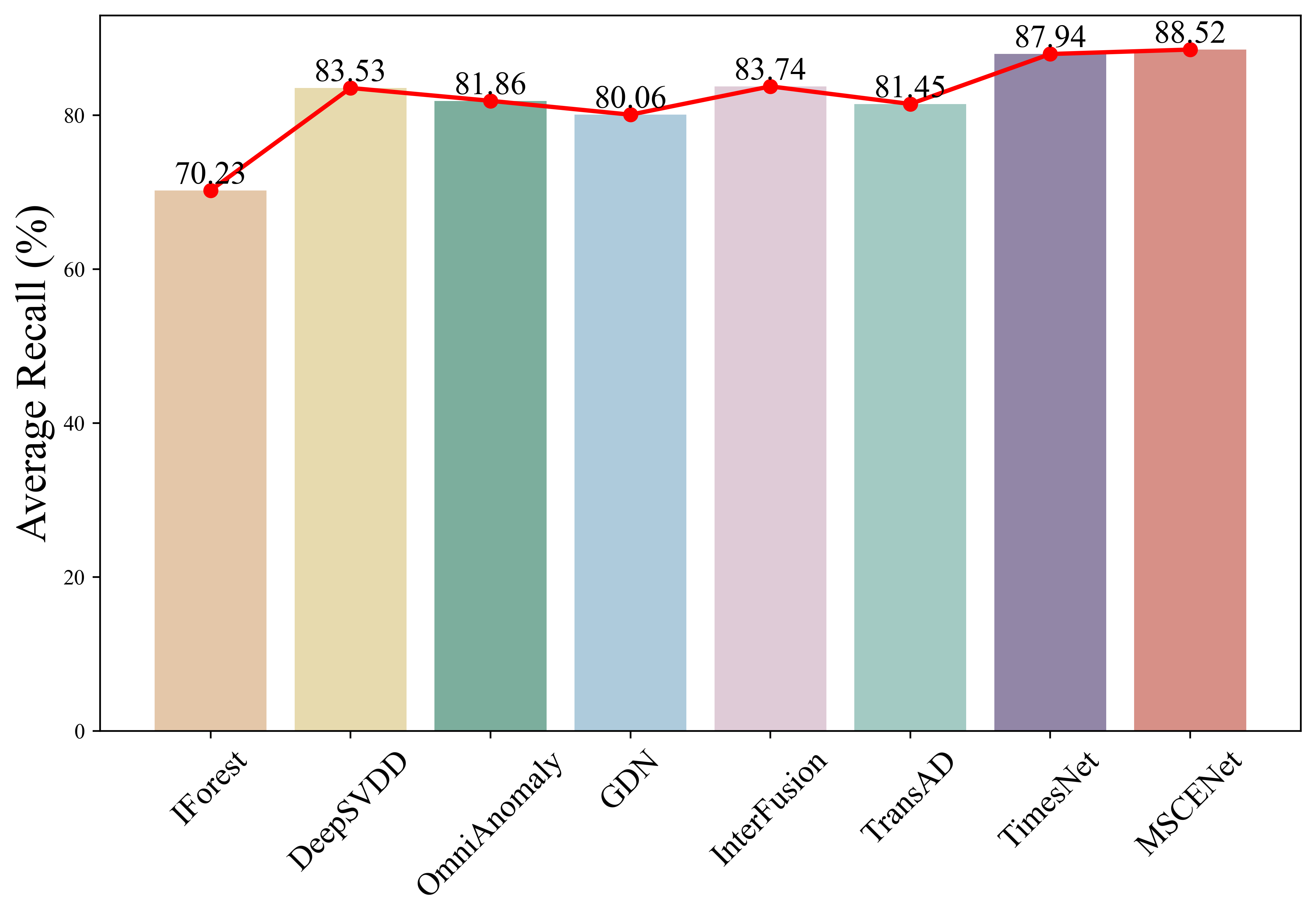}
    \\
    \makebox[0.45\textwidth]{\small (b) Average Recall Comparison}
    \\
    \includegraphics[width=0.40\textwidth]{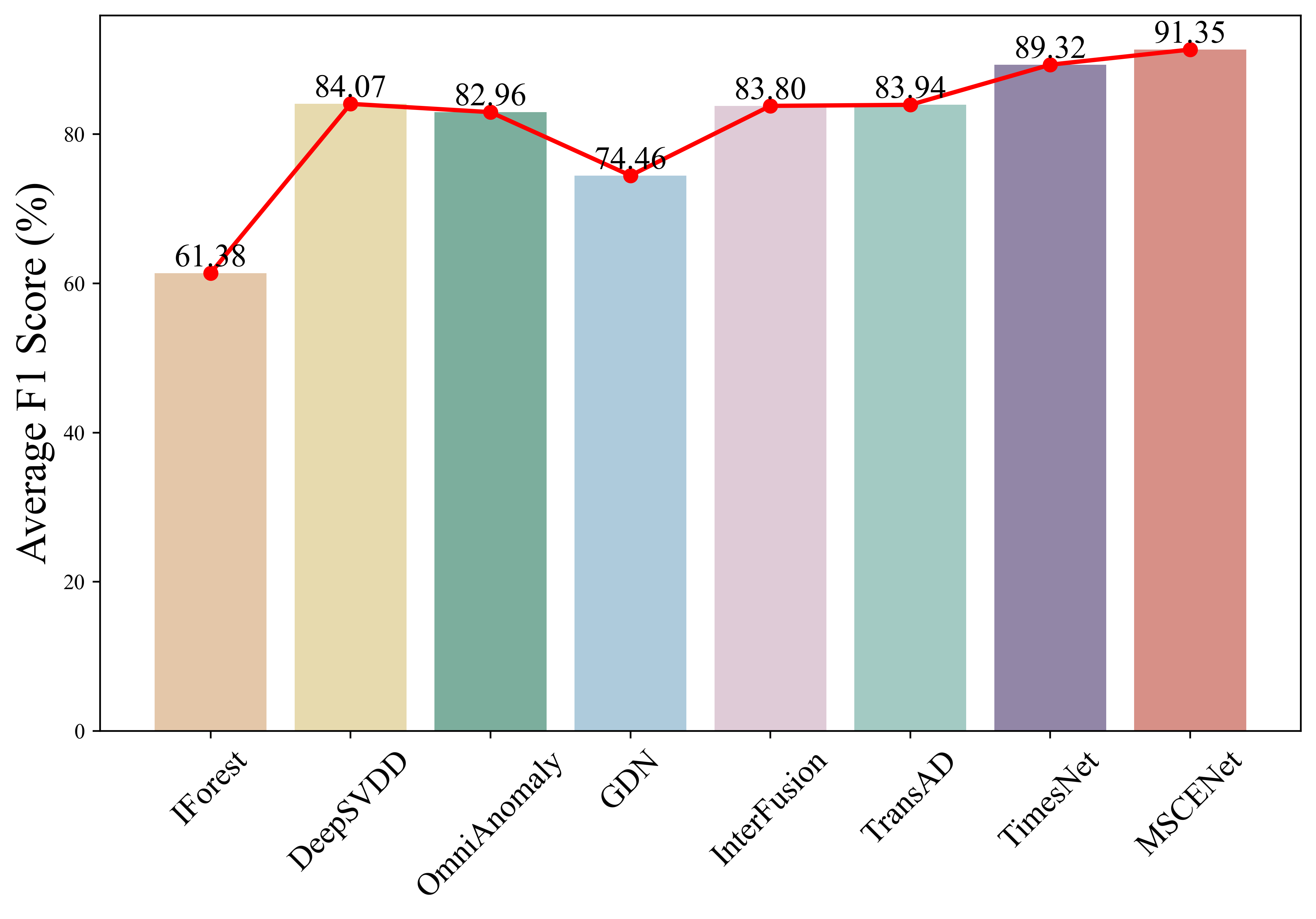}
    \\
    \makebox[0.45\textwidth]{\small (c) Average F1-score Comparison}
    \caption{Performance comparison of MSCENet and baseline methods.}
    \label{fig:average_comparison}
\end{figure}

As illustrated in Fig. \ref{fig:average_comparison} MSCENet consistently surpasses competing methods on the SMD, PSM, and SWaT datasets, achieving the highest average F1-score (91.35\%), precision (94.73\%), and recall (88.52\%), showcasing its robustness and adaptability for multivariate time series anomaly detection. In contrast to conventional approaches such as Isolation Forest, which struggle with complex temporal and spatial dependencies, MSCENet balances precision and recall effectively, reducing false positives while maintaining high accuracy. Advanced models such as Deep-SVDD, OmniAnomaly, GDN, InterFusion, TranAD and TimesNet capture some temporal dependencies but lack the flexibility for dynamic inter-series relationships and multi-scale adaptation, often resulting in missed anomalies or false positives. By integrating multi-scale temporal features and leveraging graph-based spatial learning, MSCENet captures dependencies at different scales. This flexibility allows it to handle complex datasets that feature continuously evolving dependencies, thereby achieving highly precise anomaly detection.

\subsection{Ablation Studies}
To understand the impact of individual components on the MSCENet framework, a series of ablation experiments are performed. In these experiments, we systematically removed or modified key components of the model, including the fine-grained temporal convolution module, the spatial correlation learning module, and the multi-scale gated convolution mechanism, to evaluate their influence on anomaly detection effectiveness across all datasets.

\input{table/ablation_studies}

\begin{itemize}
  \item \textbf{w/o FGTConv}: The FGTConv layer in MSCENet is crucial for modeling complex time dependencies in time series data through dilated convolutions, which extend the receptive field without additional computational cost. This allows the model to capture both immediate and extended patterns, which are key for detecting anomalies with high accuracy. Removing FGTConv leads to notable performance declines across datasets. For instance, on the SMD dataset, precision drops from 88.24\% to 88.09\% and recall from 85.71\% to 83.02\%, indicating reduced sensitivity to true anomalies. Similarly, on the PSM dataset, F1-score decreases from 97.29\% to 95.54\%, underscoring the module’s importance for detecting multi-scale temporal patterns. Theoretically, FGTConv enhances MSCENet’s ability to capture anomalies manifesting across diverse scales, which is particularly valuable in applications like industrial monitoring and finance, where both transient and gradual trends may signal critical anomalies. The observed drops in recall emphasize FGTConv’s role in strengthening MSCENet’s sensitivity to temporal anomalies, supporting robust detection across complex, multivariate time series.
  \item \textbf{w/o GConv}: The spatial correlation learning module in MSCENet is essential for capturing inter-series relationships by dynamically constructing a graph of multivariate time series data, allowing the model to detect anomalies driven by complex inter-variable interactions. Removing GConv results in consistent declines across metrics. For instance, on the SMD dataset, F1-score decreases from 86.68\% to 84.28\%, precision from 88.24\% to 87.88\%, and recall from 85.71\% to 80.97\%. Similarly, on the PSM dataset, the F1-score drops from 97.29\% to 95.70\%, with recall falling from 96.17\% to 92.93\%, indicating that MSCENet’s sensitivity to true anomalies weakens without GConv. Theoretically, GConv supports spatial dependency modeling, aligning with graph-based approaches that reflect dynamic inter-variable relationships. Practically, its presence is vital for real-world applications, such as industrial monitoring, where anomalies often result from evolving inter-series dependencies. The declines in certain metrics, particularly recall and F1-score, underscore GConv’s critical role in enhancing MSCENet’s robustness and adaptability for multivariate anomaly detection.
  \item \textbf{w/o GatedConv}: The multi-scale gated convolution in MSCENet is crucial for fusing spatial and temporal features across multiple receptive fields, enabling the model to capture dependencies at various temporal scales and adapt to diverse anomaly patterns. When GatedConv is removed, we observe consistent declines in precision, recall, and F1-score across all datasets. The F1-score on the SMD dataset falls from 86.68\% to 85.27\%, precision decreases from 88.24\% to 87.78\%, and recall drops from 85.71\% to 82.91\%, signaling a reduction in the model’s capacity to detect real anomalies and avoid false positives. Similarly, on the SWaT dataset, the F1-score drops from 90.09\% to 88.13\%, with notable declines in recall, emphasizing that GatedConv strengthens the model’s balance between sensitivity and specificity. Theoretically, GatedConv’s multi-scale design aligns with hierarchical temporal processing principles, allowing MSCENet to capture short- and long-term patterns essential for nuanced anomaly detection. Practically, it is valuable for real-world applications, such as industrial monitoring, where detecting multi-scale anomalies can prevent costly failures. Declines in recall and F1-score emphasize GatedConv’s role in capturing multi-scale temporal dependencies, critical for MSCENet’s robustness, even as precision remains stable.
\end{itemize}

\section{Conclusion}
\label{sec:conclusion}
To tackle the difficulties in identifying anomalies within multivariate time series, we have presented MSCENet, a neural network framework enhanced by multi-scale correlations. MSCENet has addressed the limitations of traditional and current deep learning methods by integrating three essential components: a fine-grained temporal convolution module, a spatial correlation learning module based on adaptive graphs, and a multi-scale gated convolution module.
Evaluation of MSCENet on real-world datasets (SMD, PSM, and SWaT) has shown that the framework has consistently outperformed other models, having achieved superior F1-scores, precision, and recall. Future research can prioritize three areas: improving the computational efficiency of the model, expanding the scope of application of the model, and exploring the practical application of MSCENet in scenarios requiring accurate anomaly detection.
\bibliographystyle{IEEEtran}

\bibliography{reference.bib}
\bibliography{reference.bib}

\vspace{-3\baselineskip}
\begin{IEEEbiography}[{\includegraphics[width=1in,height=1.25in,clip,keepaspectratio]{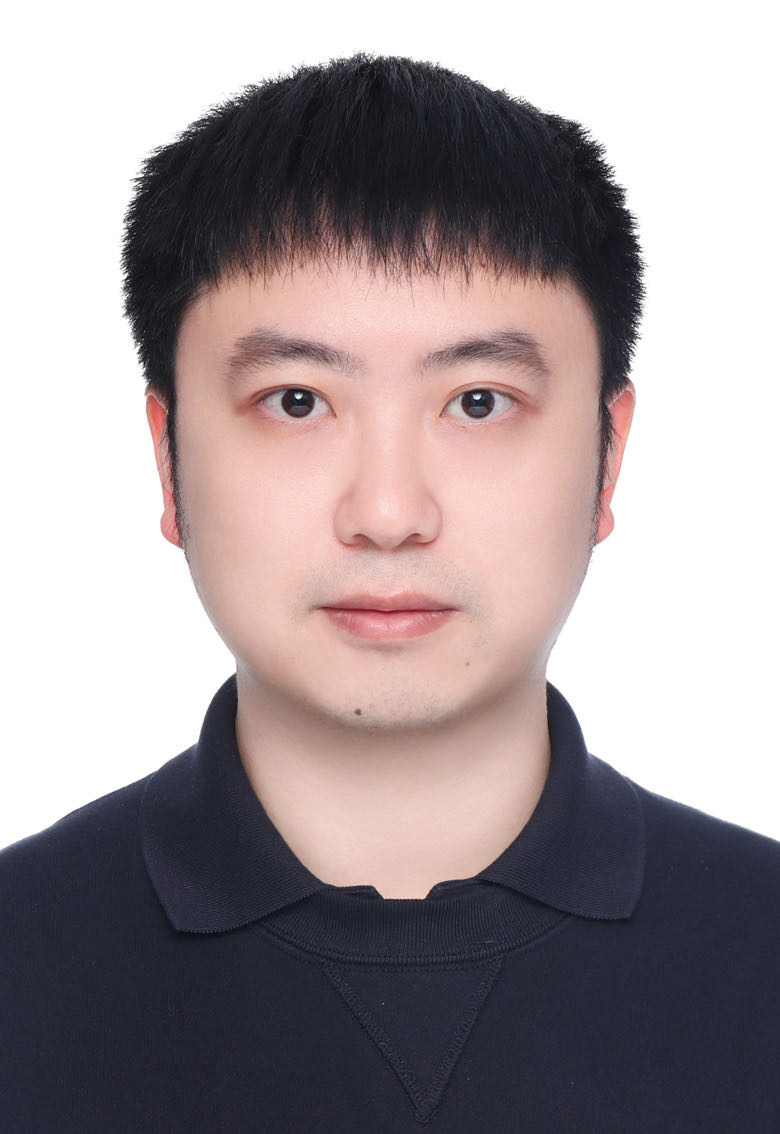}}]{Long Zhao}
	is currently pursuing the master's degree in Artificial Intelligence and Automation at the School of Electronic and Information Engineering, Tongji University, Shanghai, China. His research interests include network embedding and anomaly detection.
\end{IEEEbiography}

\vspace{-3\baselineskip} 

\begin{IEEEbiography}[{\includegraphics[width=1in,height=1.25in,clip,keepaspectratio]{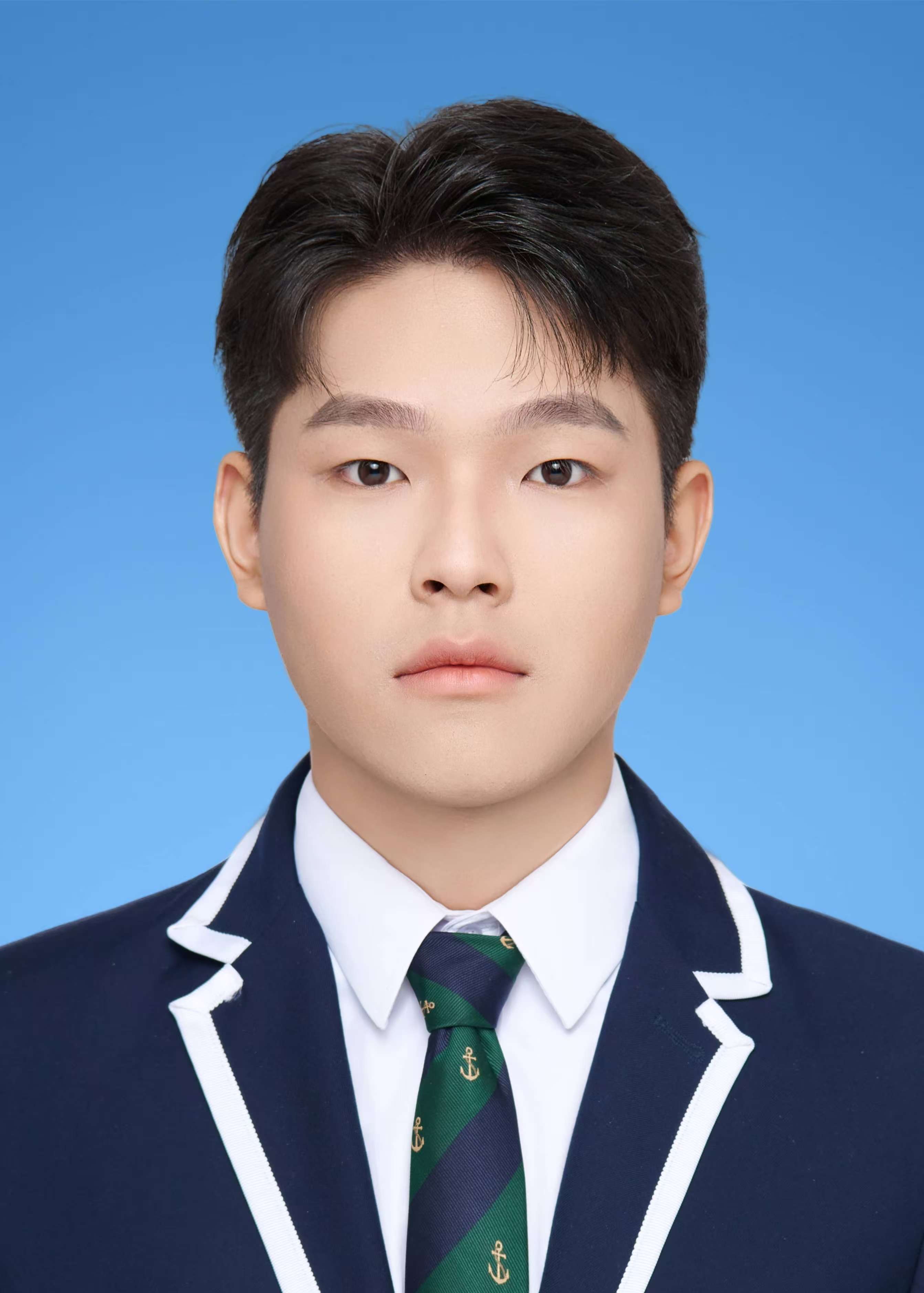}}]{Shixun Ji}
	is currently pursuing the master's degree in Artificial Intelligence and Automation at the School of Electronic and Information Engineering, Tongji University, Shanghai, China. His research interests include anomaly detection.
\end{IEEEbiography}

\vspace{-3\baselineskip} 

\begin{IEEEbiography}[{\includegraphics[width=1in,height=1.25in,clip,keepaspectratio]{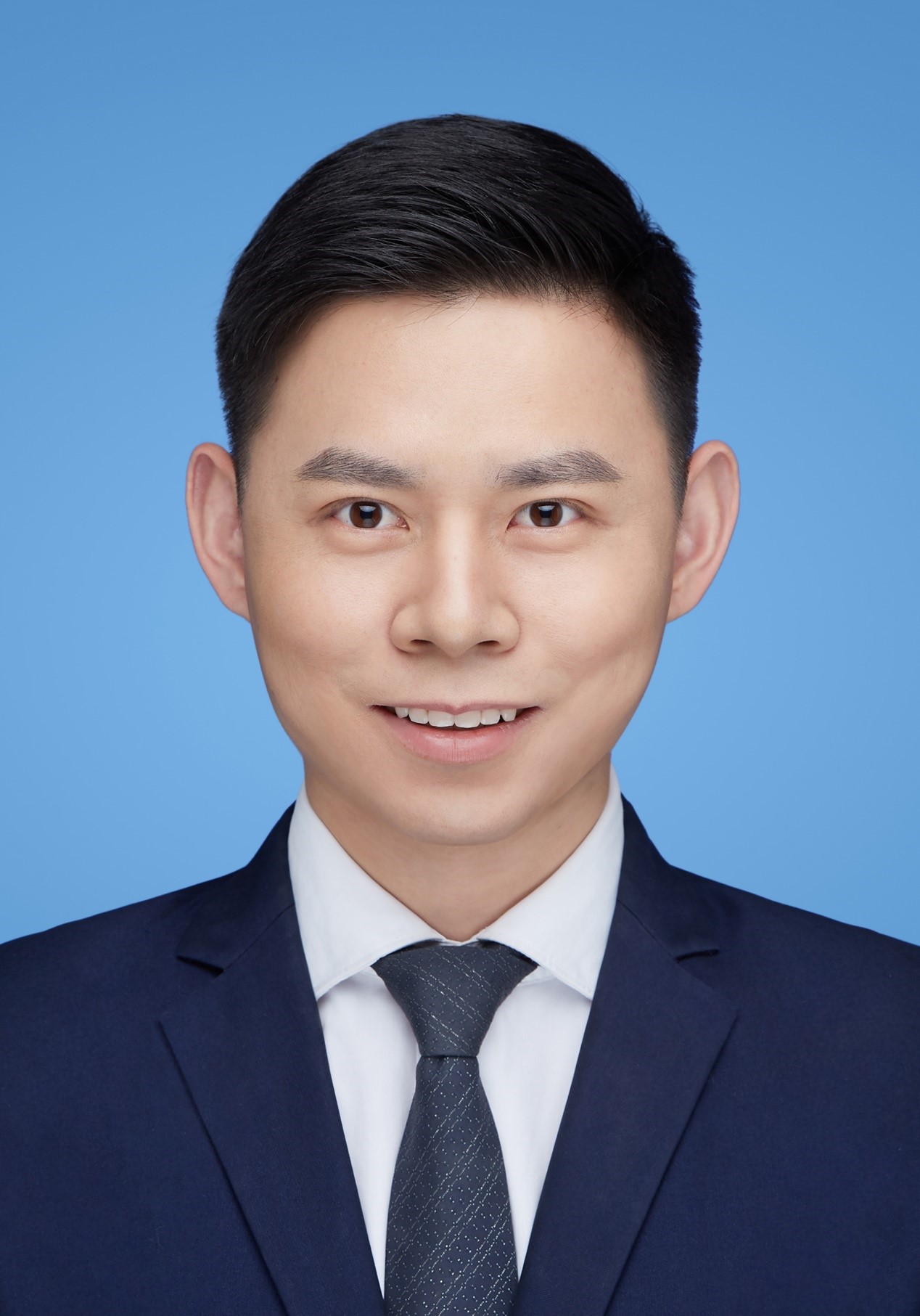}}]{Bin Cheng}
	(Member, IEEE) received the B.S. degree in mechanical engineering and automation from the School of Mechanical Engineering, University of Science and Technology, Beijing, China, in 2015, and the Ph.D. degree in dynamical systems and control from the Department of Mechanics and Engineering Science, College of Engineering, Peking University, Beijing, in 2020.
	
	He is currently an Associate Professor with the Department of Control Science and Engineering, College of Electronics and Information Engineering, Tongji University, Shanghai, China. His current research interests include cooperative control of multi agent systems, adaptive control, event-triggered control, and cooperative perception. 
\end{IEEEbiography}

\vspace{-3\baselineskip} 

\begin{IEEEbiography}
	[{\includegraphics[width=1in,height=1.25in,clip,keepaspectratio]{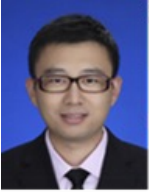}}]{Zhipeng Wang}
	(Member, IEEE) received the M.S. degree from Zhejiang University, Hangzhou, China, in 2011, and the Ph.D. degree from Tongji University, Shanghai, China, in 2015. Between 2015 and 2018, he held postdoctoral research appointments with the College of Mechanical Engineering, Tongji University. He is currently a vice professor with the College of Electronics and Information Engineering, Tongji University. His current research interests include biped robot, mechatronics and dynamics.
\end{IEEEbiography}

\vspace{-3\baselineskip} 

\begin{IEEEbiography}[{\includegraphics[width=1in,height=1.25in,clip,keepaspectratio]{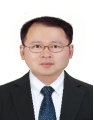}}]{Bin He}
	(Senior Member, IEEE) received the Ph.D. degree in mechanical and electronic control engineering from Zhejiang University, Hangzhou, China, in 2001, where he held postdoctoral research appointments with The State Key Lab of Fluid Power Transmission and Control, from 2001 and 2003. 
	
	He is currently a Professor with the College of Electronics and Information Engineering, Tongji University, Shanghai, China. His current research interests include intelligent robot control, biomimetic microrobots, and wireless networks.
\end{IEEEbiography}

\end{document}

%% file: table/datasets_detail.tex
\begin{table*}[htbp]
\centering
\footnotesize
\caption{Details of datasets. AR illustrates the true abnormal proportion of the dataset as a whole.}
\label{tab:benchmark-details}
\begin{tabular}{*{8}{c}}
\toprule
\textbf{Benchmarks} & \textbf{Applications} & \textbf{Dimension} & \textbf{Window} & \textbf{\#Training} & \textbf{\#Validation} & \textbf{\#Test (labeled)} & \textbf{AR (Truth)} \\
\midrule
SMD        & Server            & 38         & 100        & 566,724       & 141,681       & 708,420      & 0.042  \\
PSM        & Server            & 25         & 100        & 105,984       & 26,497        & 87,841       & 0.278  \\
SWaT       & Water             & 51         & 100        & 396,000       & 99,000        & 449,919      & 0.121  \\
\bottomrule
\end{tabular}
\end{table*}

%% file: table/baseline_comparison.tex
\begin{table*}[htbp]\centering
    \footnotesize
    \caption{Precision, Recall, and F1-score (F1) for Time Series Anomaly Detection on SMD, PSM and SWaT datasets. Higher values indicate better performance.}
    \label{tab:baseline_comparison}
    \begin{tabular}{*{10}{c}}
        \toprule
        \multicolumn{1}{c}{\textbf{Datasets}} & \multicolumn{3}{c}{\textbf{SMD}} & \multicolumn{3}{c}{\textbf{PSM}} & \multicolumn{3}{c}{\textbf{SWaT}} \\
        \cmidrule(lr){2-4} \cmidrule(lr){5-7} \cmidrule(lr){8-10}
        \multicolumn{1}{c}{\multirow{1}{*}{\textbf{Metrics}}} & \textbf{Precision} & \textbf{Recall} & \textbf{F1} & \textbf{Precision} & \textbf{Recall} & \textbf{F1} & \textbf{Precision} & \textbf{Recall} & \textbf{F1} \\
        \midrule
        \textbf{Isolation Forest} & 
        42.31 & 73.29 & 53.64 & 
        76.09 & 92.45 & 83.48 & 
        46.29 & 44.95 & 47.02  \\

        \textbf{Deep-SVDD} & 
        78.54 & 79.67 & 79.10 & 
        95.41 & 86.49 & 90.73 & 
        80.42 & 84.45 & 82.39  \\

        \textbf{OmniAnomaly} & 
        83.68 & 86.82 & 85.22 & 
        88.39 & 74.46 & 80.83 & 
        81.42 & 84.30 & 82.83  \\

        \textbf{GDN} & 
        71.70 & \textbf{99.74} & 83.42 & 
        50.47 & 70.88 & 58.96 & 
        96.97 & 69.57 & 81.01  \\

        \textbf{InterFusion} & 
        87.77 & 82.20 & 84.89 & 
        83.61 & 83.45 & 83.52 & 
        80.59 & 85.58 & 83.01  \\

        \textbf{TransAD} & 
        77.74 & 79.03 & 78.38 & 
        95.67 & 76.74 & 85.17 & 
        87.99 & 88.58 & 88.28  \\

        \textbf{TimesNet} & 
        87.94 & 81.54 & 84.62 & 
        \textbf{98.56} & 90.66 & 94.45 & 
        86.32 & \textbf{91.64} & 88.90  \\

        \textbf{Ours} & 
        \textbf{88.24} & 85.71 & \textbf{86.68} & 
        98.44 & \textbf{96.17} & \textbf{97.29} &  
        \textbf{97.53} & 83.70 & \textbf{90.09}\\
        \bottomrule
    \end{tabular}
\end{table*}

%% file: table/ablation_studies.tex
\begin{table*}[htbp]\centering
    \footnotesize
    \caption{Precision, Recall, and F1-score (F1) for Time Series Anomaly Detection on SMD, PSM and SWaT datasets. Higher values indicate better performance.}
    \label{tab:baseline_comparison}
    \begin{tabular}{*{10}{c}}
        \toprule
        \multicolumn{1}{c}{\textbf{Datasets}} & \multicolumn{3}{c}{\textbf{SMD}} & \multicolumn{3}{c}{\textbf{PSM}} & \multicolumn{3}{c}{\textbf{SWaT}} \\
        \cmidrule(lr){2-4} \cmidrule(lr){5-7} \cmidrule(lr){8-10}
        \multicolumn{1}{c}{\multirow{1}{*}{\textbf{Metrics}}} & \textbf{Precision} & \textbf{Recall} & \textbf{F1} & \textbf{Precision} & \textbf{Recall} & \textbf{F1} & \textbf{Precision} & \textbf{Recall} & \textbf{F1} \\
        \midrule
        \textbf{w/o FGTConv} & 
        88.09 & 83.02 & 85.48 & 
        98.07 & 93.14 & 95.54 &  
        98.95 & 79.10 & 87.92  \\ 

        \textbf{w/o GConv} & 
        87.88 & 80.97 & 84.28 & 
        \textbf{98.63} & 92.93 & 95.70 &  
        98.99 & 79.53 & 88.20  \\

        \textbf{w/o GatedConv} & 
        87.78 & 82.91 & 85.27 & 
        98.26 & \textbf{96.31} & 97.27 &  
        \textbf{99.09} & 79.35 & 88.13  \\

        \textbf{Ours} & 
        \textbf{88.24} & \textbf{85.71} & \textbf{86.68} & 
        98.44 & 96.17 & \textbf{97.29} &  
        97.53 & \textbf{83.70} & \textbf{90.09} \\
        \bottomrule
    \end{tabular}
\end{table*}